\documentclass[aps,prl,reprint,groupedaddress]{revtex4-2}

\usepackage{amsmath}
\usepackage{xcolor}
\usepackage{lipsum}
\usepackage{subcaption}
\usepackage{graphicx}
\usepackage{amssymb}
\usepackage[shortlabels]{enumitem}
\usepackage{bm}

\usepackage{amsthm}

\begin{document}


\title{Bifurcation and Hysteresis in Vacuum Diodes}


\author{Ashmita Panda}
\affiliation{School of Nuclear Engineering, Purdue University, West Lafayette, Indiana 47907, USA}

\author{Allen L. Garner}
\email[]{algarner@purdue.edu}
\affiliation{School of Nuclear Engineering, Purdue University, West Lafayette, Indiana 47907, USA}
\affiliation{Elmore Family School of Electrical and Computer Engineering, Purdue University, West Lafayette, Indiana 47907, USA}
\affiliation{Department of Agricultural and Biological Engineering, Purdue University, West Lafayette, Indiana 47907, USA}


\date{\today}

\begin{abstract}
    For electrons with nonzero initial velocity $u_0\ne0$ in vacuum, the space-charge-limited current density (SCLCD) is given by $J_{Jaff\acute{e}}$. For current densities $J>J_{Jaff\acute{e}}$, electron reflections occur along with virtual cathode (VC) oscillations. Past studies have exhibited hysteresis by sustaining these oscillations until  $J=J_{hys}$, often claiming $J_{hys}\approx J_{LD}<J_{Jaff\acute{e}}$, where $J_{LD}$, also called the bifurcation solution, represents the steady-state current density where the electron velocity is zero at the VC. In this study, we demonstrate that $J_{LD}$ is not a valid steady-state solution since it represents a higher charge density than $J_{Jaff\acute{e}}$, the true SCLCD. We further demonstrate that $J_{Jaff\acute{e}}$ also corresponds to the true, mathematical bifurcation solution. Using particle-in-cell simulations across various gap distances and voltages, we demonstrate that $J_{hys}$ consistently differs from $J_{LD}$ and can be represented by a simple semi-empirical relationship as a function of $J_{Jaff\acute{e}}/J_{CL}$, where $J_{CL}$ is the SCLCD in vacuum for $u_0=0$. 
\end{abstract}


\maketitle


\textbf{Introduction.-} 
Characterizing space-charge contributions to electron emission is critical for numerous applications, including high-power microwaves \citep{HPM_Book,HighPowerMicrowaves1,HighPowerMicrowaves2,HighPowerMicrowaves3}, vacuum nanotransistors \citep{Nanotransistors1,ZhangBallistics,100YearsofDiodes,YLiReviewofElectronEmission}, and thermionic energy converters \citep{campbell2021progress, khalid2016review, thermionic2026}. The space-charge-limited current density (SCLCD) corresponds to the maximum steady-state current density permissible in a diode \citep{ZhangBallistics,CLMultiDimensions}. For electrons emitted with initial velocity $u_0=0$ from an infinitesimally small region on an infinitely large planar cathode, the SCLCD is given by the Child-Langmuir law as \citep{Child,Langmuir,CLMultiDimensions}
\begin{equation}
    J_{CL}=\dfrac{4 \epsilon_0}{9}\sqrt{\dfrac{2e}{m}}\dfrac{V^{3/2}}{D^2}\ ,\label{eq:1}
\end{equation}
where $V$ is the voltage of the anode with a grounded cathode, $D$ is the gap distance, $m$ is the electron mass, $e$ is the electron charge, and $\epsilon_0$ is the permittivity of free space. 

Equation (\ref{eq:1}) can be derived using charge-free capacitance \citep{CLVacuumCapacitance}, variational calculus to extremize the current \citep{DarrGarnerNonplanar}, and electron transit time \citep{breen2024collisional, birdsallbridges_book}. Similarly, (\ref{eq:1}) may be extended to nonplanar and multidimensional geometries using uniform SCLCD \citep{Luginsland2DPIC,lau2001simple,CL2D}, variational calculus \citep{DarrGarnerNonplanar}, conformal mapping \citep{harshaconformalmapping}, point transformations \citep{LiePoint_Harsha,HarshaRelativistic}, charge-free capacitance \citep{wrightcap2026}, and charge-free electric field \citep{wright2026calculating}. While many of these approaches have focused on $u_0=0$, calculations using uniform SCLCD \citep{ZhuJAPpaper} and point transformations \citep{LiePoint_Harsha} have also considered $u_0\neq 0$ .

For nonmagnetic diodes with $u_0 \ne 0$ (the focus of this Letter), two solutions for the SCLCD exist in the literature: the Jaff\'e solution \citep{Jaffe} and the Liu-Dougal (LD) solution \citep{LiuDougalSolution}. They share a common mathematical form of 
\begin{equation}
    J_{SCL}=J_{CL}\left[ \left(\dfrac{mu_0^2}{2eV} \right)^\xi  + \left( 1+\dfrac{mu_0^2}{2eV} \right)^\xi \right]^\alpha\, , \label{eq:2}
\end{equation}
where $(\alpha,\xi)=(3,1/2)$ yields the Jaff\'e solution ($J_{SCL}=J_{Jaff\acute{e}}$), which represents the true SCLCD \cite{AkimovBifurcation,Lafleur_2020,LiePoint_Harsha}, and $(\alpha,\xi)=(2,3/4)$ yields the LD solution ($J_{SCL}=J_{LD}$), which corresponds to the theoretical limit that allows multi-streaming electron trajectories \cite{LiuDougalSolution}. 

For $u_0=0$, the electric field at the physical cathode is zero ($E_C=0$) \citep{ZhangBallistics,100YearsofDiodes,DarrGarnerNonplanar}. This implies that the electron acceleration goes to zero, leading to electron accumulation at the cathode. However, for $u_0\neq 0$, the electric field goes to zero between the cathode and anode, creating a potential minimum that is referred to as a virtual cathode (VC) \citep{100YearsofDiodes,jiang2025tutorial}. Electron accumulation occurs at the VC for $u_0\neq 0$. 

The Jaff\'e solution ($J_{Jaff\acute{e}}$) is derived by identifying the critical value of the potential minimum (i.e., VC formation) corresponding to the maximum current \citep{Lafleur_2020,LiePoint_Harsha}. The LD solution ($J_{LD}$) is derived by further enforcing zero electron velocity at the VC ($u=0$) \citep{LiuDougalSolution}. 

The concept of hysteresis behavior in diodes, where the VC oscillations that arise for $J>J_{SCL}$ are sustained even when $J<J_{SCL}$, has been long-studied \citep{harshahysteresis2D,gartstein1998hysteresis,liu2024two,czarczynski1966hysteresis,AkimovBifurcation}. Specifically, the VC oscillations are sustained until reaching a current density $J_{hys}<J_{SCL}$. Currently, $J_{hys}$ is treated interchangeably with $J_{LD}$ (i.e., $J_{hys}=J_{LD}$). This can be traced back to the work of Akimov \textit{et al.} \citep{AkimovBifurcation} and Puri \textit{et al.} \citep{Puripaper}, who provided a physical interpretation for $J_{LD}$. They argue that $J_{LD}$ corresponds to the physical condition when the solution set of the diode ``bifurcates" such that electrons may follow one of three possible electron trajectories for all $J>J_{LD}$: two without any particle reflections (one stable and one unstable) and one with particle reflections (with an oscillatory virtual cathode). Additionally, for $J<J_{LD}$, there is only a stable electron trajectory with no reflections. Thus, they refer to $J_{LD}$ as the ``bifurcation" solution that corresponds to $J_{hys}$, although the true mathematical bifurcation \citep{golubitsky1984bifurcation,edelstein2005mathematical, strogatz2024nonlinear} occurs at the space-charge-limit. They further conclude that the diode does not support any steady state solution for $J>J_{Jaff\acute{e}}$, meaning that $J_{Jaff\acute{e}}$ is the true SCLCD \citep{AkimovBifurcation}.

While several theoretical arguments attempt to relate $J_{hys}$ to $J_{LD}$, there are few simulation studies \citep{harshahysteresis2D,kuznetsov_akimov_numerical2004}. Hysteresis is often portrayed by plotting $E_C$ as a function of $J$ \citep{AkimovBifurcation,Puripaper}. Some papers alternatively plot VC location as a function of $J$ \citep{Liu2DVChysteresis, liu2024two}. Such plots cannot be made for the full range of $J$ values due to the lack of VC formation for all $J<J_{Jaff\acute{e}}$. We contend that there could be other parameters that better demonstrate the stored memory of the diode for a more useful hysteresis curve. 

This Letter addresses three issues. First, we use particle-in-cell (PIC) simulations to characterize hysteresis for different diode conditions to identify the physical quantities that yield a hysteresis curve for all diodes. Second, we demonstrate the disagreement between $J_{LD}$ from (\ref{eq:2}) and $J_{hys}$ from PIC simulations. Finally, we propose a universal semi-empirical solution for $J_{hys}$ based on PIC.

\textbf{Diode Equations.-}
Theoretically, at steady state, charge continuity gives
\begin{equation}
    J=e\,n(x)\,u(x), \label{eq:3}
\end{equation}
where $n(x)$ is the electron density and $u(x)$ is the electron velocity along the $x$-axis with the cathode at $x=0$ and anode at $x=D$. Integrating $n(x)$ across the gap gives the total charge density as
\begin{equation}
    Q_{tot}=J\, T\, , \label{eq:4}
\end{equation}
where $T$ is the electron transit time.  

Using Poisson's equation and momentum conservation and applying Llewellyn's transformation to transform from position to time \cite{birdsallbridges_book}  gives the electric field $E(t)$ and electron velocity $u(t)$ as functions of time $t$ as
\begin{equation}
    E(t)=-\dfrac{Jt}{\epsilon_0}+E_C\, , \label{eq:5}
   \end{equation}
and

\begin{equation}
    u(t)=\dfrac{e J}{2m\epsilon_0}\,t^2-\dfrac{e E_C}{m}\, t+u_0\, , \label{eq:6}
\end{equation}
where $E_C$ is the electric field at the physical cathode, i.e., $E(t=0)=E_C$. Integrating $u(t)$ with respect to $t$ yields the electron position $x(t)$ as 
\begin{equation}
    x(t)= \dfrac{eJ}{6\epsilon_0 m}\,t^3 - \dfrac{eE_C}{2m}\, t^2+u_0 t\, . \label{eq:7}
\end{equation}
Finally, defining initial velocity $u(t=0)=u_0$, final velocity $u(t=T)=u_f=\sqrt{u_0^2+2eV/m}$, initial position $x(t=0)=0$, and final position $x(t=T)=D$ gives 
\begin{equation}
    E_C=\dfrac{JT}{3\epsilon_0}+\dfrac{2mu_0}{eT}-\dfrac{2Dm}{eT^2}, \label{eq:8}
\end{equation}
and \cite{AkimovBifurcation}
\begin{equation}
    \dfrac{e}{6\epsilon_0 m}J \, T^3-(u_0+u_f)\,T+2D=0\, \label{eq:9}.
\end{equation}
To simplify these equations, we define $\beta_0=\sqrt{mu_0^2/(2eV)}$, $\beta_f=\sqrt{m u_f^2/(2eV)}$ \cite{LiePoint_Harsha}, $\tilde{T}=T\, D^{-1} \sqrt{eV/m}$, and $\tilde{J}=J/J_{CL}$. Note that $\beta_f=\sqrt{1+\beta^2_0}$ from conservation of energy. Using these non-dimensional parameters, we can rewrite (\ref{eq:9}) as
\begin{equation}
    f(\tilde{T},\tilde{J})=\tilde{J} \, \tilde{T}^3-27(\beta_0+\beta_f)\,\tilde{T}/2 +27/\sqrt{2}=0 \,. \label{eq:10}
\end{equation}
We can solve $f(\tilde{T},\tilde{J})$ to obtain $\tilde{T}$ for a given $\tilde{J}$. Rewriting $\tilde{J}_{Jaff\acute{e}}$ and $\tilde{J}_{LD}$ \citep{Lafleur_2020,Halpern, LiePoint_Harsha} using this non-dimensionalization as
\begin{equation}
    \tilde{J}_{Jaff\acute{e}}=\left(\beta_0+\beta_f\right)^3\, , \quad \tilde{J}_{LD}=\left(\beta_0^{3/2}+\beta_f^{3/2}\right)^2\, . \label{eq:11}
\end{equation}

\textbf{Bifurcation Solution.-}
Mathematically, a true bifurcation solution requires a qualitative change in the structure of the solution set \citep{golubitsky1984bifurcation}. In terms of $f(\tilde{T},\tilde{J})$, with $\tilde{T}$ as the variable and $\tilde{J}$ as the bifurcation parameter, assume a bifurcation point $(\tilde{T}_*,\tilde{J}_*)$ exists. Then, $f(\tilde{T},\tilde{J})$ must satisfy two conditions. First, $f(\tilde{T},\tilde{J})$ must have a tangential root at $(\tilde{T}_*,\tilde{J}_*)$, meaning that $f'(\tilde{T}_*,\tilde{J}_*)=0$ and $f(\tilde{T}_*,\tilde{J}_*)=0$, where the prime denotes differentiation with respect to $\tilde{T}$, because $\tilde{J}$ is treated as a constant \citep{golubitsky1984bifurcation}. Second, the number of roots of $f(\tilde{T},\tilde{J})$ for $\tilde{J}>\tilde{J}_*$ and $\tilde{J}<\tilde{J}_*$ must differ \citep{golubitsky1984bifurcation}.
\begin{figure}
    \centering
    \includegraphics[width=\linewidth]{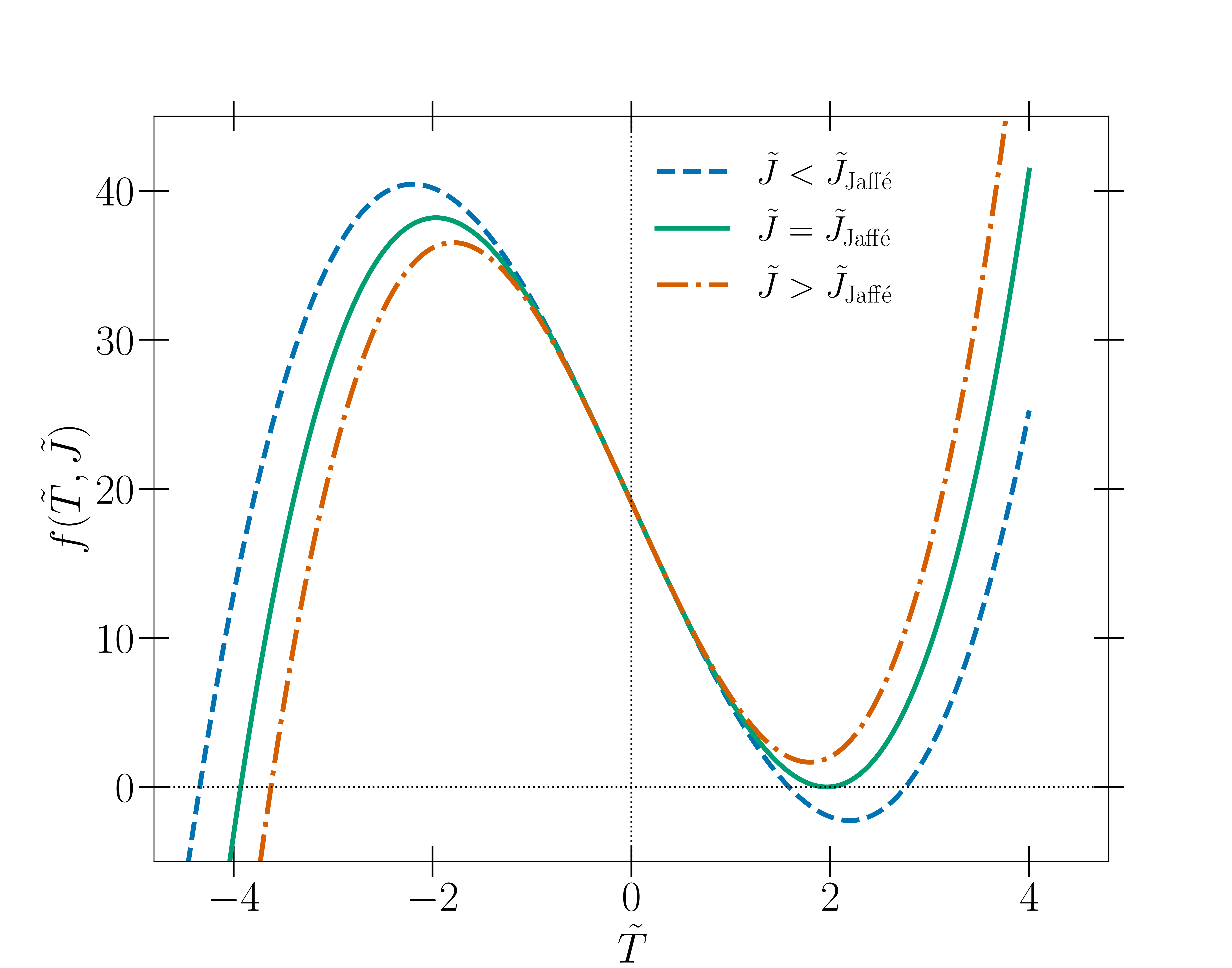}
    \caption{The roots of $f(\tilde{T},\tilde{J})$ for $\beta_0=0.077$, which corresponds to $\tilde{J}_{Jaff\acute{e}}=1.259$, plotted for $\tilde{J}=0.8\,\tilde{J}_{Jaff\acute{e}}\,$; $\tilde{J}=\tilde{J}_{Jaff\acute{e}}\,$; and $\tilde{J}=1.2\, \tilde{J}_{Jaff\acute{e}}\,$. When $J=J_{Jaff\acute{e}}$, $f(\tilde{T},\tilde{J})$ has one root, indicating bifurcation.}
    \label{fig:fJTplot}
\end{figure}

Solving for the tangential root condition yields $\tilde{J}_*=(\beta_0+\beta_f)^3=\tilde{J}_{Jaff\acute{e}}$. Additionally, as discussed by Akimov \textit{et al}. \cite{AkimovBifurcation}, Fig. \ref{fig:fJTplot} shows that for a chosen $\beta_0$, the number of roots for $f(\tilde{T},\tilde{J})$ changes as $\tilde{J}$ crosses $\tilde{J}_*=\tilde{J}_{Jaff\acute{e}}$. For $\tilde{J}<\tilde{J}_{Jaff\acute{e}}$, there exist two positive roots, $\tilde{T}_1$ and $\tilde{T}_2$, such that $\tilde{T}_1<\tilde{T}_2$. For $\tilde{J}=\tilde{J}_{Jaff\acute{e}}$, the two positive solutions converge to a single root, $\tilde{T}=\tilde{T}_{Jaff\acute{e}}$, while for $\tilde{J}>\tilde{J}_{Jaff\acute{e}}$, there are no positive roots. Hence, $(\tilde{T}_{Jaff\acute{e}},\tilde{J}_{Jaff\acute{e}})$ corresponds to the true mathematical bifurcation solution for the vacuum diode, \textit{not} the LD solution.

We ignore the negative root since negative transit times are not physically relevant. Further analysis of the two positive roots verifies that $\tilde{T}_1$ increases, whereas $\tilde{T}_2$ decreases as $\tilde{J}$ increases. Both roots approach $\tilde{T}_{Jaff\acute{e}}$ as $\tilde{J}$ approaches $\tilde{J}_{Jaff\acute{e}}$. Thus, $\tilde{T}_1$ and $\tilde{T}_2$ form two separate solution branches that converge at the bifurcation point. The first branch spans $0<\tilde{T}\leq \tilde{T}_{Jaff\acute{e}}$ while the second branch spans $\tilde{T}_{Jaff\acute{e}}\leq \tilde{T}$. This raises a question concerning the physical relevance of the LD solution and the solution branch where it resides.  
\begin{figure}
    \centering
    \includegraphics[width=\linewidth]{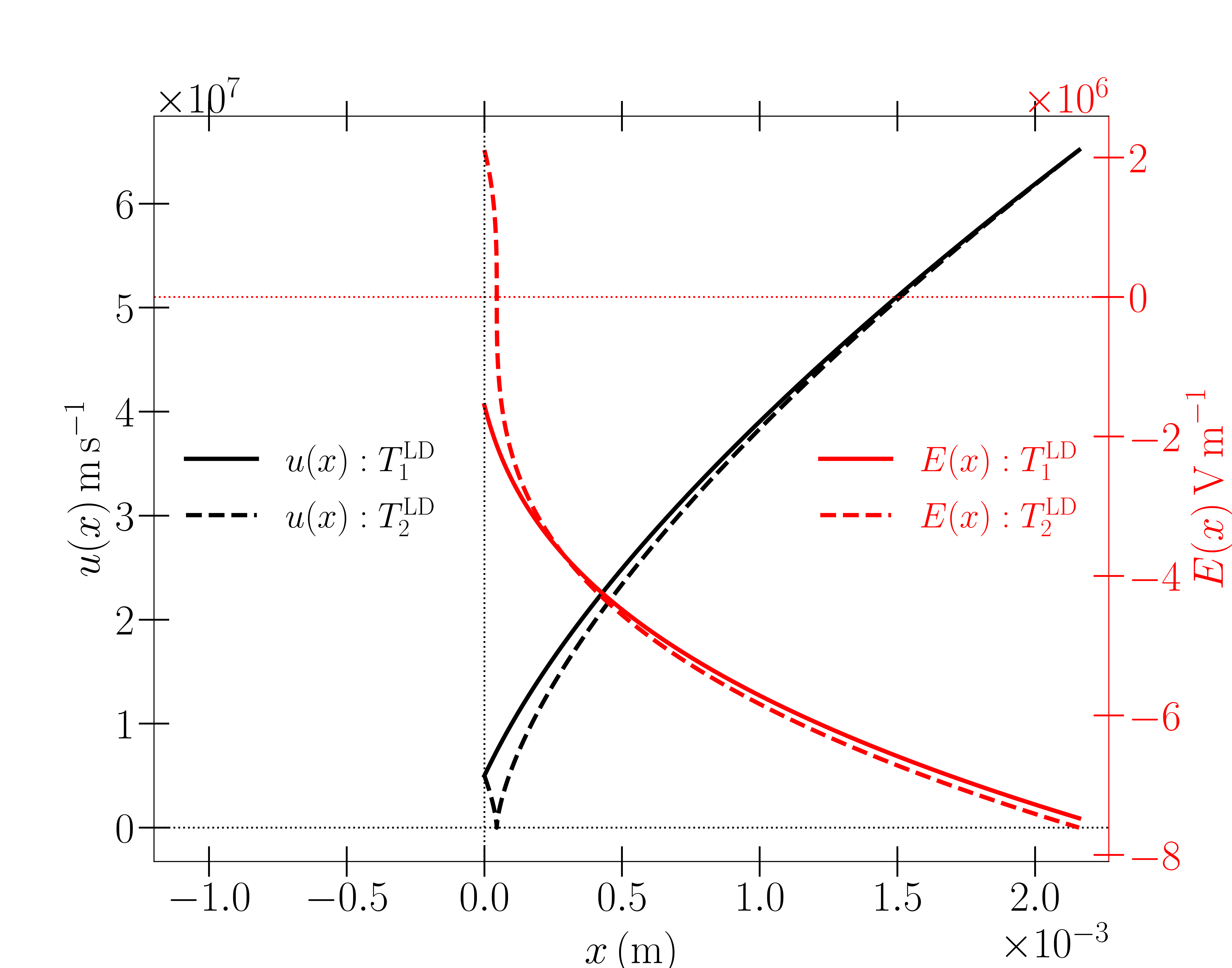}
    \caption{Electron trajectory and electric field inside a gap with separation distance $D=0.00216$ m, cathode voltage of $V=-12000$ V with a grounded anode, and $u_0=5\times10^6$ m s$^{-1}$.}
    \label{fig:trajectories}
\end{figure}

The LD solution corresponds to the combination of $(\tilde{T}_{LD},\tilde{J}_{LD})$ at which $u(x=x_{VC})=0$, which is the minimum possible electron velocity inside the gap, at the VC (defined by $E(x=x_{VC})=0$). Since $\tilde{J}_{LD}<\tilde{J}_{Jaff\acute{e}}$ \cite{AkimovBifurcation,halpern2022coordinate}, (\ref{eq:10}) has two possible roots at $\tilde{J}=\tilde{J}_{LD}$. Both roots, $\tilde{T}^{LD}_1$ and $\tilde{T}^{LD}_2$, defined such that $\tilde{T}^{LD}_1<\tilde{T}^{LD}_2$, can be identified numerically using root finding algorithms. Figure \ref{fig:trajectories} shows the electron trajectory and electric field in the gap for the two different roots for a chosen $\beta_0$ (corresponding to a chosen $u_0$). The LD condition is only fulfilled by $\tilde{T}^{LD}_2$, implying the LD solution lies on the second branch. 

Furthermore, analysis of current densities $\tilde{J}<\tilde{J}_{LD}$ on the second solution branch, corresponding to $\tilde{T}_2>\tilde{T}^{LD}_2$, show that electron trajectories start turning around [i.e., $u(t)<0$ in the gap]. This is forbidden by the steady state theory \cite{garner1998collapse}. Hence, $\tilde{J}_{LD}$ forms the bound of the second solution branch that spans $\tilde{T}_{Jaff\acute{e}}\leq \tilde{T} \leq \tilde{T}_{LD}$.


\textbf{Hysteresis curve.-}
Now that we have demonstrated that the LD solution does not correspond to mathematical bifurcation, we next characterize hysteresis in a diode. The plot of magnetic flux $B$ as a function of magnetizing force $H$ for ferromagnets is the most well-known hysteresis curve in physics \cite{chikazumi1997physics}. Hysteresis curves for diodes focus on current-voltage plots, primarily for solar cells \cite{solarcell1}. Essentially, hysteresis curves demonstrate the internal ``memory" of the system \cite{moree2023review}. Thus, we propose that an equivalent hysteresis curve for a diode would plot the total charge density $Q_{tot}$.
\begin{figure}
    \centering
    \includegraphics[width=\linewidth]{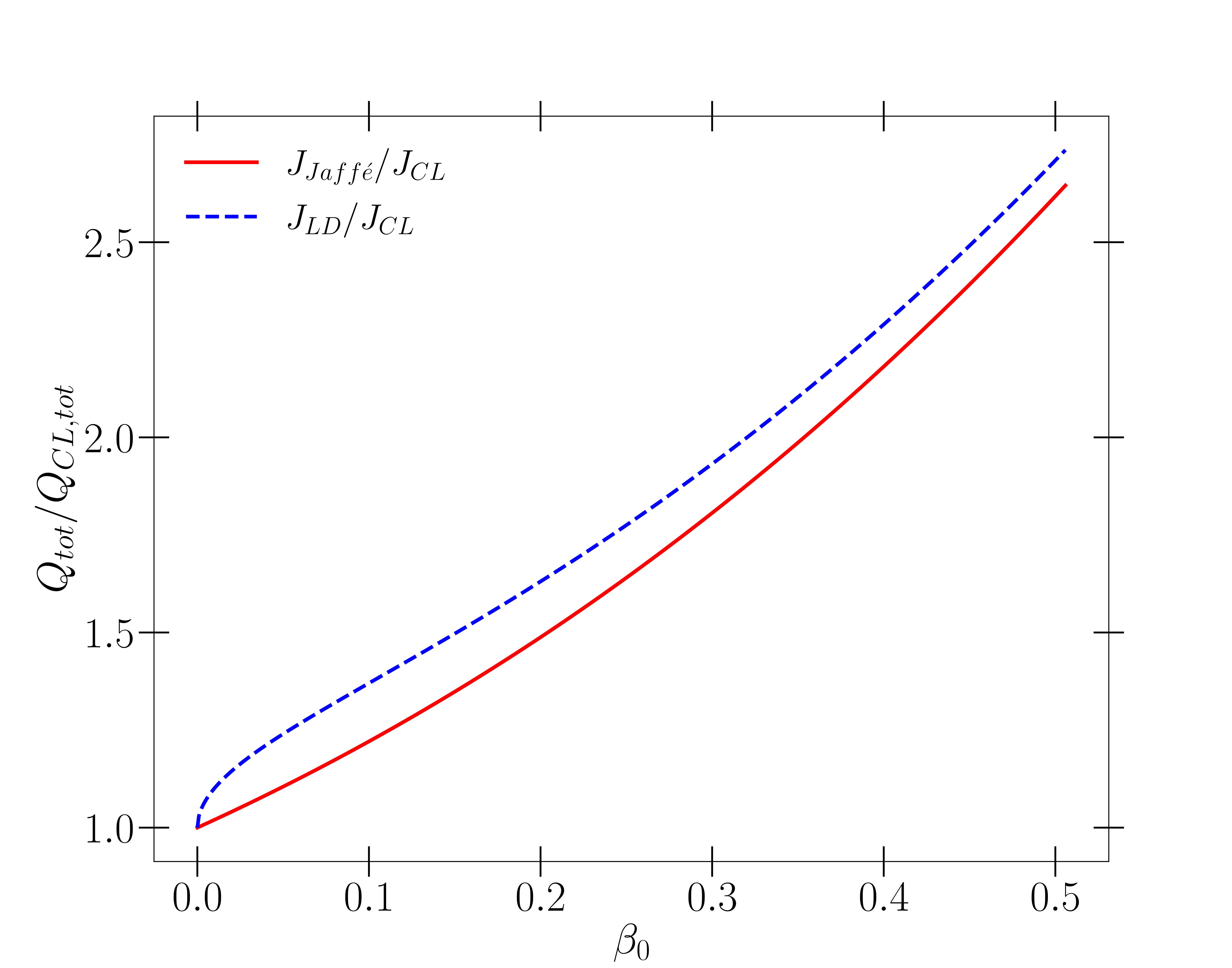}
    \caption{Total charge density $Q_{tot}$ in a gap for $J=J_{LD}$ or $J=J_{Jaff\acute{e}}$ normalized to the charge density in a space-charge limited gap, $Q_{CL,tot}$, with normalized initial electron energy $\beta_0=0$ as a function of $\beta_0$ .}
    \label{fig:qratiobeta0}
\end{figure}

We can define $\tilde{Q}_{tot}=Q_{tot}/Q_{CL,tot}$ as the normalized total charge density in a space-charge-limited (SCL) diode with $\beta_0\ne 0$ scaled to the charge in a SCL diode with $\beta_0=0$. Figure \ref{fig:qratiobeta0} shows that for all $\beta_0>0$, $\tilde{Q}_{tot,LD}>\tilde{Q}_{tot,Jaff\acute{e}}$. This indicates a major issue with the LD solution, which is derived from steady state theory. Since $\tilde{J}_{Jaff\acute{e}}$ is the true SCLCD in a vacuum diode \citep{AkimovBifurcation, Lafleur_2020, LiePoint_Harsha}, adding charge once $\tilde{J}=\tilde{J}_{Jaff\acute{e}}$ causes electron reflections, making it the steady-state limit. Thus, any solution corresponding to $\tilde{Q}_{tot}>\tilde{Q}_{tot,Jaff\acute{e}}$ is inadmissible as a steady state solution, which makes $\tilde{J}_{LD}$ \textit{inherently inconsistent with the steady state theory} used to derive it.

We next compare the theoretical calculations of $\tilde{Q}_{tot}$, including the LD solution, to PIC simulations. Akimov \textit{et al.} argue that $\tilde{J}_{LD}$ corresponds to the minimum current density that will sustain VC oscillations once they start (i.e., $\tilde{J}_{hys}=\tilde{J}_{LD}$) \cite{AkimovBifurcation}. We use the one-dimensional in space, three-dimensional in velocity electrostatic PIC code, XPDP1 \cite{XPDP1}, to assess the validity of this statement. We set $D=0.00216 \, \text{m}$, cathode area $A=0.003122 \, \text{m}^2$, and hold the cathode at $V=-12 000 \, \text{V}$ while the anode is grounded, based on the U.S. Navy Aegis radar system \cite{PeggyThesis}. We set the time step as $5\times 10^{-13} \ $s and the number of cells to $500$ to ensure numerical stability \cite{revolinsky2024two} and the number of physical particles per macroparticle to $1\times 10^8$ to maintain approximately $10^3$ to $10^5$ macroparticles. 

For a fixed injection velocity, the simulation is run for different injection current densities, $\tilde{J}_{inj}$. After each run, the simulated diode state, including the electron positions, electron velocities, electric field, and electric potential on the spatial grid, along with the exact timestep, is stored in a dump file. In the next run, with a different $\tilde{J}_{inj}$, the diode state is restored from the dump file. This mimics the act of slowly changing $\tilde{J}_{inj}$ by choosing appropriately small steps of $\Delta \tilde{J}_{inj}$. After each run, the time-averaged electron distribution is also used to determine the total charge stored in the gap. 
\begin{figure}
    \centering
    \includegraphics[width=\linewidth]{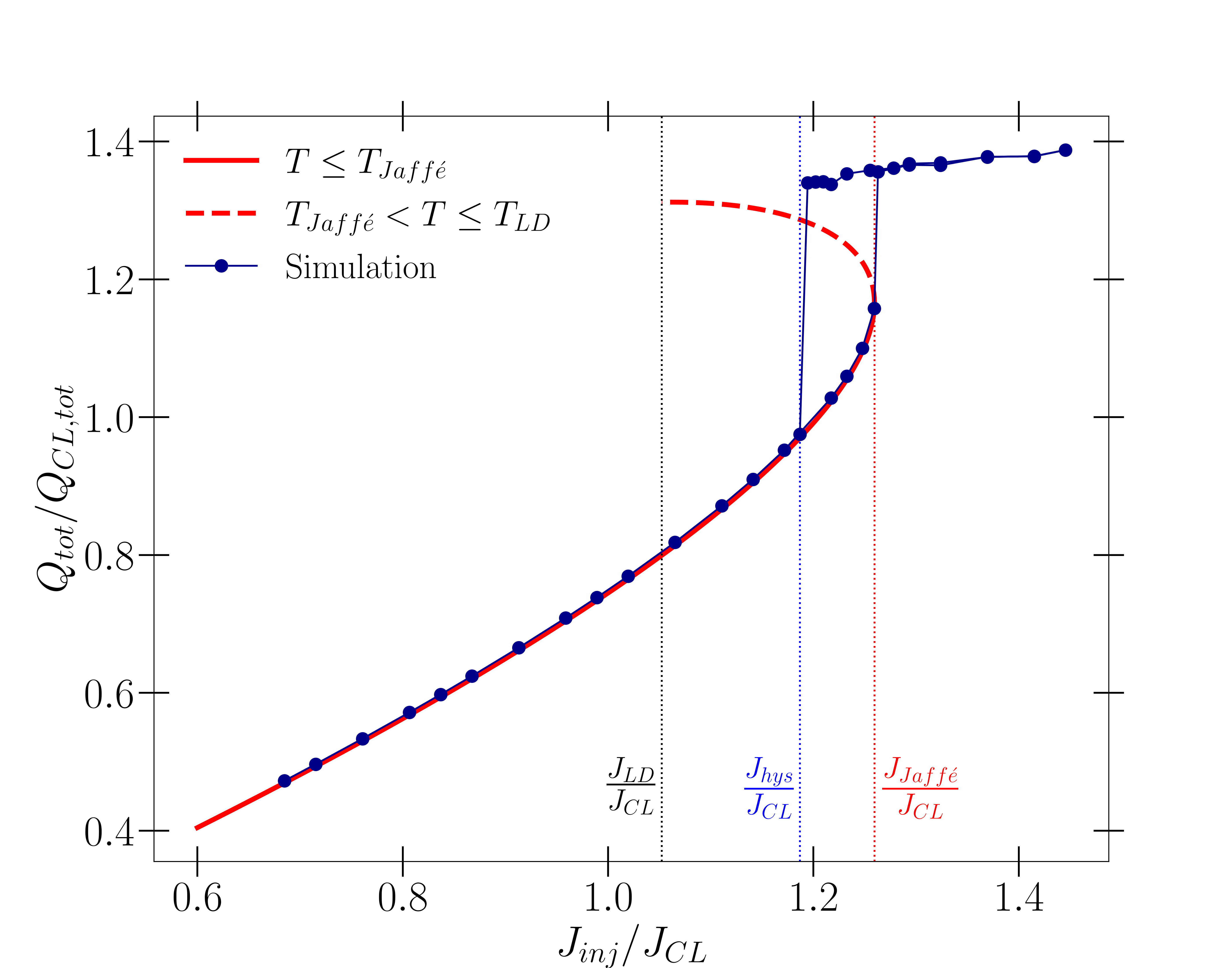}
    \caption{Total charge density $Q_{tot}$ in the gap, for $u_0=5\times10^6\, $m s$^{-1}$, with $D=0.00216$ m, cathode voltage $V=-12000$ V and a grounded anode, scaled to the total charge in the gap for a space-charge-limited gap with $u_0=0$, $Q_{CL,tot}$, as a function of the injected current density $J_{inj}$ scaled to the Child-Langmuir law, $J_{CL}$. The simulated results follow the theoretical results up to the space-charge-limit at $J_{inj}=J_{Jaff\acute{e}}$ before spiking. The current density returns to the stable curve at $J_{inj}=J_{hys}\neq J_{LD}$.} 
    \label{fig:qratiovsJratio}
\end{figure}
Figure \ref{fig:qratiovsJratio} shows that the simulated and theoretical $\tilde{Q}_{tot}$ agree for a chosen $\beta_0$ until $\tilde{J}_{inj}=\tilde{J}_{Jaff\acute{e}}$. The charge density in the gap increases abruptly at $\tilde{J}_{inj}=\tilde{J}_{Jaff\acute{e}}$, indicating the onset of VC oscillations, and then increases approximately linearly with further increasing $\tilde{J}_{inj}$. As $\tilde{J}_{inj}$ is reduced, the charge density returns along the same path it followed with increasing $\tilde{J}_{inj}$ until $\tilde{J}_{inj}=\tilde{J}_{Jaff\acute{e}}$. As $\tilde{J}_{inj}$ is reduced below $\tilde{J}_{Jaff\acute{e}}$, the charge density does not follow either the $\tilde{T}\leq \tilde{T}_{Jaff\acute{e}}$ branch or the $\tilde{T}_{Jaff\acute{e}}<\tilde{T}\leq \tilde{T}_{LD}$ branch. Instead, VC oscillations continue to be sustained by the diode. At some $\tilde{J}_{hys}\neq \tilde{J}_{LD}$, the total charge rejoins the theoretical $\tilde{T}\leq \tilde{T}_{Jaff\acute{e}}$ branch, indicating the collapse of VC oscillations.

\begin{figure}
    \centering
    \includegraphics[width=\linewidth]{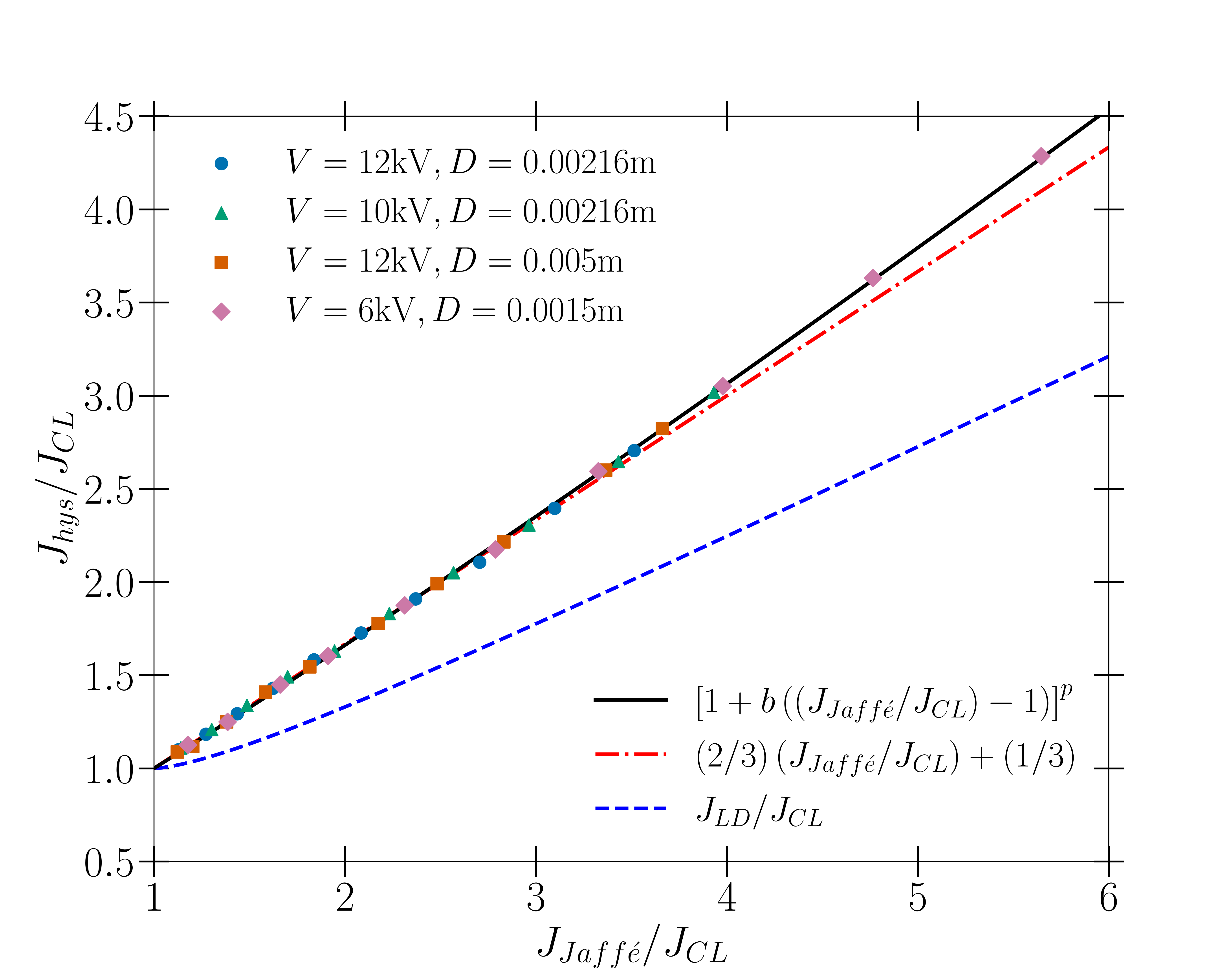}
    \caption{Hysteresis current density $J_{hys}$ as a function of the space-charge-limited current density $J_{Jaff\acute{e}}$ for nonzero initial velocity $u_0$ chosen between $2.5\times 10^6 \text{ m}\, \text{s}^{-1}$ to $2.8\times 10^7 \text{ m}\, \text{s}^{-1}$, both normalized to the Child-Langmuir current density $J_{CL}$. Fitting the complete dataset to (\ref{eq:12}) yields  $b=0.5753$ and $p=1.1165$ with individual $R^2$ given in Table \ref{tab:1}.}
    \label{fig:jhysratio}
\end{figure}

\textbf{Hysteresis current density.-}
We repeated the simulations for different values of $u_0$, $V$, and $D$ to determine the dependence of $\tilde{J}_{hys}$ on the parameters, and compare it to $\tilde{J}_{LD}$. We chose $u_0$ between $2.5\times10^6\text{ m}\,\text{s}^{-1} $ and $2.8\times10^7\text{ m}\,\text{s}^{-1} $ to ensure the electrons remain non-relativistic. Figure \ref{fig:jhysratio}, which shows $\tilde{J}_{hys}$ as a function of $\tilde{J}_{Jaff\acute{e}}$, demonstrates the disagreement between the simulated $\tilde{J}_{hys}$ and theoretical $\tilde{J}_{LD}$. Thus, $\tilde{J}_{hys}$ is \textit{not} related to $\tilde{J}_{LD}$, which itself would only be valid in steady state if it were physically achievable. Thus, we conclude that $\tilde{J}_{LD}$ is both physically invalid and practically inaccurate for estimating hysteresis in a diode.

We instead propose an ansatz for $\tilde{J}_{hys}$ as
\begin{equation}
    \tilde{J}_{hys}=\left[1+b\,\left(\tilde{J}_{Jaff\acute{e}}-1\right)\right]^p\, . \label{eq:12}
\end{equation}
Equation (\ref{eq:12}) recovers $\tilde{J}_{hys}=1$ for $\tilde{J}_{Jaff\acute{e}}=1$, which corresponds to $\beta_0=0$. We determine $b$ and $p$ by fitting the simulation results to (\ref{eq:12}). We obtain $b=0.5753$ and $p=1.1165$ with an overall $R^2=0.99953$ for the complete dataset collected in Fig. \ref{fig:jhysratio}.

For sufficiently small $\tilde{J}_{Jaff\acute{e}}$, we can apply the binomial expansion to (\ref{eq:12}) to rewrite it as a linear equation. For $\tilde{J}_{Jaff\acute{e}}\lesssim 2.5$, (\ref{eq:12}) can be approximated as
\begin{equation}
    \tilde{J}_{hys}\approx\dfrac{2}{3}\, \tilde{J}_{Jaff\acute{e}}+\dfrac{1}{3}\, . \label{eq:13}
\end{equation}
\begin{table}[h]
    \centering
    \begin{tabular}{|c|c|c|c|}
        \hline
         \multicolumn{2}{|c|}{\textbf{Diode Parameters}}& \textbf{(\ref{eq:12})} & \textbf{(\ref{eq:13})} \\
         \cline{1-2}
         \textbf{Voltage (kV)} & \textbf{Gap distance (m)} & $\bm{R^2}$ & $\bm{R^2}$ \\
         \hline
         12 & 0.00216 & 0.9984 & 0.9989 \\
         \hline
         10 & 0.00216 & 0.9993 & 0.9984 \\
         \hline
         12 & 0.005 & 0.9996 & 0.9989 \\
         \hline
         6 & 0.0015 & 0.9999 & 0.9957 \\
         \hline
    \end{tabular}
    \caption{Coefficient of determination, $R^2$, for fitting  simulation results for various diode parameters to (\ref{eq:12}) with $b=0.5753$ and $p=1.1165$, and (\ref{eq:13}).}
    \label{tab:1}
\end{table}

For $\tilde{J}_{Jaff\acute{e}}>2.5$, the contributions of the higher order terms in the binomial expansion require using the full equation from (\ref{eq:12}). Table \ref{tab:1} summarizes $R^2$ for the fitting to (\ref{eq:12}) with $b=0.5753$ and $p=1.1165$, and (\ref{eq:13}).

\textbf{Conclusion.-}
Starting from the equation for current density as a function of electron transit time in a one-dimensional vacuum diode with non-zero electron velocity, we used the mathematical definition of bifurcation to recover Jaff\'e's solution for SCLCD \cite{Jaffe} as the true bifurcation solution for the steady state theory. This reinforced the validity of this solution as the true SCLCD \citep{AkimovBifurcation,Lafleur_2020,LiePoint_Harsha}, corresponding to the highest possible steady state solution. 

We also determined that there exist two possible transit time solutions for $J<J_{Jaff\acute{e}}$. The two solutions lie on separate branches that coincide only at the Jaff\'e limit. For increasing $J$, the transit time increases for the lower branch and decreases for the higher branch. Combining the electron trajectory behavior with the definition of the LD solution showed that the LD solution lies on the higher branch, implying that the electrons reside longer in the diode for the LD solution compared to the Jaff\'e solution. This means that achieving the LD solution would require the gap to hold more charge than the Jaff\'e solution. Since this is inadmissible, the LD solution is inherently inconsistent with steady state theory. Using XPDP1, we demonstrated that the total charge stored inside the diode agrees closely with the theoretically predicted values for the lower transit time branch, while the higher branch containing the LD solution does not materialize. While this was expected, as the LD solution is often described as an ``unstable" physical state where the diode does not operate, we still expected the VC oscillations to be sustained by the diode until $J_{LD}$ \cite{AkimovBifurcation}. PIC simulations further demonstrated that VC oscillations were not sustained until $J_{LD}$. Instead, we provide an alternate equation to determine the current density $J_{hys}$ at which the diode returns to steady state. 

In conclusion, diode hysteresis is a complex phenomenon that depends on diode geometry, initial conditions, and the total charge in the gap. An accurate equation for this behavior cannot be derived from steady state theory assuming constant current density and forward flow of electrons across the diode. The best way to characterize this behavior is from simulations. Our simple equation based on simulation results provides a valuable guide for predicting this behavior that may ultimately be extended to more realistic geometries {\cite{wright2026calculating}}.
\section*{Acknowledgments}
This material is based upon work supported by the Air
Force Office of Scientific Research under Award No. FA9550-22-1-0434. The views expressed are those of the authors and do not reflect the official policy or position of the Department of War or the U.S. Government.
\bibliography{DiodeHysteresisBibliography}

\end{document}